\documentclass[12pt]{article}

\usepackage{setspace}
\usepackage{epsfig}
\usepackage{amssymb}
\def\be{\begin{equation}}
\def\ee{\end{equation}}
\def\ba{\begin{array}}
\def\ea{\end{array}}
\def\beqn{\begin{eqnarray}}
\def\eeqn{\end{eqnarray}}

\def\bt{\begin{tabular}}
\def\et{\end{tabular}}
\def\bc{\begin{center}}
\def\ec{\end{center}}
\begin{document}

\title{$~~~~~~~~$Exploring the parameter space of texture 4 zero quark mass matrices}

\author{Rohit Verma$^1$, Gulsheen Ahuja$^2$, Neelu Mahajan$^2$, \\ Monika Randhawa$^3$, Manmohan Gupta$^2$\\
\\
 {$^1$ \it Rayat Institute of Engineering and Information Technology, Ropar, India.} \\
{$^2$ \it Department of Physics, Centre of Advanced Study, P.U.,
 Chandigarh, India.}\\
 {$^3$ \it University Institute of Engineering and Technology, P.U., Chandigarh, India.} \\
 {\it Email: mmgupta@pu.ac.in}}

 \maketitle
 \begin{abstract}
We have attempted to extend the parameter space of the elements of
the texture 4 zero Hermitian quark mass matrices, to include the
case of `weak hierarchy' amongst them along with the usually
considered `strong hierarchy' case. This has been carried out by
giving wide variation to the hierarchy defining parameters $D_U$
and $D_D$, having implications for the structural features of the
mass matrices. We find that not only the weakly hierarchical mass
matrices are able to reproduce the strongly hierarchical mixing
angles but also both the phases having their origin in the mass
matrices have to be non zero to achieve compatibility of these
matrices with recent quark mixing data. Further noting the
difference between the exclusive and inclusive values of $V_{ub}$,
we have carried out separate analyses corresponding to these.
 \end{abstract}

 \section{Introduction}
In the last few years, several important developments have taken
place in the context of phenomenology of Cabibbo-Kobayashi-Maskawa
(CKM) matrix \cite{ckm}, both from theoretical as well as
experimental point of view. In this context, it may be noted that
texture specific mass matrices seem to be very helpful in
understanding the pattern of quark mixings and CP violation
\cite{9912358, groupquarksmm}. Likewise, in the leptonic sector
also texture zero mass matrices \cite{0307359}-\cite{leptex} have
proved to be useful in explaining the pattern of neutrino masses
and mixings, some of which have been recently measured with a good
deal of accuracy. In particular, the Fritzsch-like texture four
zero quark mass matrices are quite successful in reconciling the
strong hierarchy of quark masses and the smallness of flavor
mixing angles \cite{9912358, groupquarksmm}. Similarly, in the
case of neutrinos, the texture four zero mass matrices are able to
accommodate the neutrino oscillation data representing large
mixing angles quite well \cite{0307359}-\cite{leptex}. Further,
apart from being related to the Nearest Neighbor Interaction (NNI)
form of the mass matrices through weak basis rotations \cite{nni},
these texture 4 zero mass matrices are known to be compatible with
specific models of GUTs, e.g., SO(10) \cite{9912358, 0307359,
bando, fuku} and these could be obtained using considerations of
Abelian family symmetries \cite{abelian}. Furthermore, in case we
have to consider quark-lepton unification \cite{qlepuni}, then it
becomes interesting to note that the corresponding Fritzsch-like
texture 4 zero neutrino mass matrices have been shown to be seesaw
invariant \cite{seesawinv}. Also, it may be added that the
structure of these matrices is very much in agreement with the
hypothesis of natural mass matrices advocated by \cite{naturalmm}.

It may be mentioned that in the case of quark mass matrices,
usually the elements are assumed to follow `strong hierarchy',
whereas there is no such compulsion for the leptonic mass
matrices. Therefore, in case we have to invoke quark-lepton
unification \cite{qlepuni}, it becomes interesting to examine
whether `weakly hierarchical' quark mass matrices are able to
reproduce the mixing data which involves strongly hierarchical
parameters. This is all the more important as the texture 4 zero
mass matrices perhaps provide the simplest parallel structure for
quark and lepton mass matrices which are compatible with the low
energy data.

Realizing the importance of Fritzsch-like Hermitian texture 4 zero
mass matrices in the context of quarks, as emphasized above,  a
few years back Xing and Zhang \cite{xingzhang} have attempted to
find the parameter space of the elements of these mass matrices.
Their analysis has provided good deal of information regarding the
space available to various parameters as well as have provided
valuable insight into the `structural features' of texture 4 zero
mass matrices. In this context, it may be noted that the hierarchy
of the elements of the mass matrices is largely governed by the
(2,2) element of the matrix. In their analysis, attempt has been
made to go somewhat beyond the minimal values of this element,
corresponding to the `strong hierarchy' case, however in case we
have to consider the `weak hierarchy' case as well then there
seems a further need to consider a still larger range for this
element. Further, their analysis has also given valuable clues
about the phase structure of the mass matrices, in particular for
the strong hierarchy case they conclude that only one of the two
phase parameters plays a dominant role. Since, the phases of the
mass matrices play a crucial role in giving information about the
CP violating phase of the CKM matrix, therefore it would be
interesting to find out the ranges of both the phases in the
context of strong as well as weak hierarchy of the elements of the
mass matrices.

It may be noted that in the last few years there have been
considerable improvements in the measurement of CKM parameters and
light quark masses. However, at present, even after recent
updating by various groups \cite{pdg08}-\cite{hfag}, the situation
regarding the element $V_{ub}$ is not as clear as for the other
CKM matrix elements. For example, as per PDG 2008 \cite{pdg08} its
exclusive and inclusive values respectively are
$(3.5^{+0.6}_{-0.5}) {10^{-3}}$ and $(4.12 \pm 0.43 ){10^{-3}}$.
It may be noted that although the difference between the exclusive
and inclusive values of $V_{ub}$ is not statistically significant,
however recent unitarity based analyses \cite{ouruni, oursin2b} as
well as results from a global fit \cite{utfit} emphasize exclusive
value of $V_{ub}$. Therefore, it becomes important to examine
separately the implications of exclusive and inclusive values of
$V_{ub}$ for the phenomenology of quark mass matrices.

The purpose of the present work is to update and broaden the scope
of the analysis carried out by Xing and Zhang \cite{xingzhang} as
well as to examine the implications of recent precision
measurements on the structural features of texture 4 zero mass
matrices for exclusive and inclusive $V_{ub}$ separately. Taking
clue from the quark-lepton symmetry it would also be desirable to
explore the parameter space of the elements of the mass matrices,
considering not only the usual `strong hierarchy' amongst them but
also for the `weak hierarchy' case. In view of the more precise
information regarding CP violating parameters, it would be
interesting to find out the ranges of both the phases having their
origin in the mass matrices which are compatible with the quark
mixing data. Further, for the sake of completeness it would also
be desirable to construct the CKM mixing matrix as well as to
evaluate the Jarlskog's rephasing invariant parameter $J$ and the
CP violating phase $\delta$.

The detailed plan of the paper is as follows. In Section
(\ref{form}), we detail the essentials of the formalism regarding
the texture specific mass matrices. Inputs used in the present
analysis and the methodology of the calculations have been given
in Section (\ref{inputs}). The discussion of the results have been
presented in Section (\ref{cal}). Finally, Section (\ref{summ})
summarizes our conclusions.

\section{Formalism\label{form}}
To begin with, we define the modified Fritzsch-like matrices,
e.g., \be M_i = \left( \ba {ccc} 0 & A_i & 0 \\ A_i^{*} & D_i &
B_i
\\
            0 & B_i^{*} & C_i \ea \right), \qquad i=U,D\,, \label{uniq} \ee
$M_U$ and $M_D$, respectively corresponding to the mass matrix in
the up sector and the down sector. It may be noted that each of
the above matrix is texture 2 zero type with $A_{i}
=|A_{i}|e^{i\alpha_{i}}$ and $B_{i} = |B_{i}|e^{i\beta_{i}}$. The
various relations between the elements of the mass matrices $A_i ,
B_i, C_i, D_i$ essentially correspond to the structural features
of the mass matrices including their hierarchies.

In the absence of any standard definition in the literature for
`weak' and `strong' hierarchy of the elements of the mass
matrices, for the purpose of present work we consider these as
follows. As is usual the element $|A_i|$ takes a value much
smaller than the other three elements of the mass matrix which can
assume different relations amongst each other, defining different
hierarchies. For example, in case $D_i
< |B_{i}| < C_i$ it would lead to a strongly hierarchical
mass matrix whereas a weaker hierarchy of the mass matrix implies
$D_{i} \lesssim |B_i| \lesssim C_i$. It may also be added that for
the purpose of numerical work, one can conveniently take the ratio
$D_i/C_i \sim 0.01$ characterizing strong hierarchy whereas
$D_i/C_i \gtrsim 0.2$ implying weak hierarchy. This can be
understood by expressing these parameters in terms of the quark
masses, in particular $D_U/C_U \sim 0.01$ implies $C_U \sim m_t$
and $D_D/C_D \sim 0.01$ leads to $C_D \sim m_b$.

To facilitate diagonalization, the complex mass matrix $M_i$ $(i =
U, D)$ can be expressed as
\be
M_i= Q_i M_i^r P_i \,  \label{mk} \ee or  \be M_i^r= Q_i^{\dagger}
M_i P_i^{\dagger}\,, \label{mkr} \ee where $M_i^r$ is a real
symmetric matrix with real eigenvalues and $Q_i$ and $P_i$ are
diagonal phase matrices. The matrix $M_i^r$ can be diagonalized by
the orthogonal transformation, e.g., \be M_i^{\rm diag} = O_i^T
M_i^{r} O_i
 \,,   \label{o1}\ee
where \be M_i^{\rm diag} = {\rm diag}(m_1,\,-m_2,\,m_3)\,, \ee the
subscripts 1, 2 and 3 referring respectively to $u,\, c$ and $t$
for the $U$ sector as well as $d,\,s$ and $b$ for the $D$ sector.
Using the invariants, tr$M_i^r$, tr ${M_i^r}^2$ and det$M_i^r$,
the values of the elements of the mass matrices $A_{i}$, $B_{i}$
and $C_{i}$, in terms of the free parameter $D_{i}$ and the quark
masses are given as
 \beqn
  C_i& = &(m_1-m_2+m_3-D_i)\,, \\
   |A_i| &=&(m_1 m_2 m_3/C_i)^{1/2}\,, \\
 |B_i| &= &
 [(m_3-m_2-D_i)(m_3+m_1-D_i)(m_2-m_1+D_i)/C_i]^{1/2}\,.
\label{elements} \eeqn

The exact diagonalizing transformation $O_i$ is expressed as
 \be O_i
= \left( {\renewcommand{\arraystretch}{1.7}
 \ba{ccc}
  \pm {\sqrt \frac{m_2 m_3 (C_i-m_1)}{(m_3-m_1)(m_2+m_1)C_i} } &
   \pm  {\sqrt \frac{m_1 m_3 (C_i+m_2)}{C_i (m_2+m_1) (m_3+m_2)}} &
  \pm{\sqrt \frac{m_1 m_2 (m_3-C_i)}{C_i (m_3+m_2)(m_3-m_1)}}\\
 \pm{\sqrt \frac{m_1 (C_i-m_1)}{(m_3-m_1)(m_2+m_1)} } &
 \mp{\sqrt \frac{m_2 (C_i+m_2)}{(m_3+m_2)(m_2+m_1)} }&
 \pm{\sqrt \frac{m_3(m_3-C_i)}{(m_3+m_2)(m_3-m_1)} } \\
 \mp{\sqrt \frac{m_1 (m_3-C_i)(C_i+m_2)}{C_i(m_3-m_1)(m_2+m_1)} } &
 \pm{\sqrt \frac{m_2 (C_i-m_1) (m_3-C_i)}{C_i
(m_3+m_2)(m_2+m_1)} } &
  \pm{\sqrt \frac{m_3 (C_i-m_1)(C_i+m_2)}{C_i
(m_3+m_2)(m_3-m_1)}}  \ea} \right). \label{ou} \ee \vskip 0.5cm

It may be noted that while finding the diagonalizing
transformation $O_i$, one has the freedom to choose several
equivalent possibilities of phases. Similarly, while normalizing
the diagonalized matrix to quark masses, one again has the freedom
to choose the phases for the quark masses. This is due to the fact
that the diagonalizing transformations of $M_U$ and $M_D$ occur in
a particular manner in the weak charge current interactions of
quarks to give the CKM mixing matrix. As is usual, we have chosen
the phase of $m_2$ to be negative facilitating the diagonalization
process as well as the construction of the CKM matrix. This is one
of the possibilities considered by Xing and Zhang
\cite{xingzhang}, in particular it corresponds to their $(\eta_u,
\eta_d)=(-1,-1)$. The other possibilities considered by them are
related and are all equivalent as well, these only redefine the
phases $\phi_1$ and $\phi_2$ which in any case are arbitrary. For
the present work, we have chosen the possibility, \be O_i= \left(
\ba{ccc} ~~O_i(11)& ~~O_i(12)& ~O_i(13)
\\
 ~~O_i(21)& -O_i(22)& ~O_i(23)\\
     -O_i(31) & ~~O_i(32) & ~O_i(33) \ea \right). \ee

The CKM mixing matrix $V_{\rm CKM}$ which measures the non-trivial
mismatch between diagonalizations of $M_U$ and $M_D$ can be
obtained using $O_{U(D)}$ through the relation
\be
V_{\rm CKM}= O_U^T (P_U P_D^{\dagger}) O_D. \ee Explicitly, the
elements of the CKM mixing matrix can be expressed as
\be
V_{l m} = O_{1 l}^U O_{1 m}^D e^{-i \phi_1} + O_{2 l}^U O_{2 m}^D
 + O_{3 l}^U O_{3 m}^D e^{i \phi_2},
\label{vckmelement} \ee
 where the subscripts $l$ and $m$ run respectively over $u,\, c$, $t$  and $d,\,s$,
 $b$ with $\phi_1 =  \alpha_U- \alpha_D$, $\phi_2= \beta_U-
 \beta_D$.

\section{Inputs used and calculations\label{inputs}}
Before discussing the results of our analysis, we would first like
to briefly mention the inputs used for carrying out the
calculations. We have adopted the following ranges of quark masses
\cite{xinginput} at the $M_z$ energy scale, e.g.,
 \beqn ~~~~~~m_u=1.27^{+0.5}_{-0.42}\, {\rm MeV},~~~~~~~m_d=2.90^{+1.24}_{-1.19}\, {\rm
MeV},~~~~~~~ m_s=55^{+16}_{-15}\, {\rm MeV},~~~~~~~~\nonumber\\
~~~~~~~m_c=0.619 \pm 0.084\, {\rm GeV},~~ m_b=2.89 \pm 0.09\, {\rm
GeV},~~ m_t=171.7 \pm 3.0\, {\rm GeV}. ~~~~~~~~\label{qmasses}
\eeqn The light quark masses $m_u$, $m_d$ and $m_s$ have been
further constrained using the following mass ratios given by
\cite{leut} \beqn ~~~~~~m_u / m_d =0.553 \pm
0.043,\,~~~~~~~~~~~~~m_s / m_d =18.9 \pm
 0.8.
 \label{ratios} \eeqn

Further, we have given full variation to the phases $\phi_1$ and
$\phi_2$, the parameters $D_U$ and $D_D$ have been given wide
variation in conformity with the hierarchy of the elements of the
mass matrices e.g., $D_i < C_i$ for $i=U, D$. The extended range
of these parameters allows one to carry out the calculations for
the case of weak hierarchy of the elements of the mass matrices as
well. Also, it needs to be mentioned that the present range of
$D_U$ and $D_D$ is much wider than the one considered by Xing and
Zhang \cite{xingzhang}. In particular, we have considered $D_i/C_i
\sim 0.05-0.8$, whereas Xing and Zhang have emphasized $D_i/C_i
\sim 0.1$ which corresponds to the case of strong hierarchy
amongst the elements of the mass matrices. Furthermore, we have
imposed the following constraints due to the latest PDG 2008
values \cite{pdg08}, \beqn |V_{us}|=0.2255 \pm 0.0019
,\;~~~~~~~~~~~~~~~~|V_{cb}|=(41.2 \pm 1.1) 10^{-3},~~~~~~~~~
\nonumber\\ V_{ub}({\rm excl.})= 0.0035\pm 0.0006,\;~~~~~~~~~
V_{ub}({\rm incl.})= 0.00412\pm 0.00036, \nonumber\\{\rm
sin}\,2\beta = 0.681 \pm
0.025.~~~~~~~~~~~~~~~~~~~~~~~~~~~~~~~~~~~~~~~~~~~~~~~~~~~~~~~~~~~\label{ckmvalues}
\eeqn It may be noted that the calculations have been carried out
separately for both exclusive and inclusive values of $|V_{ub}|$.

\section{Results and discussion\label{cal}}
In view of the fact that one of the aim of the present analysis is
to update as well as to extend the analysis of Xing and Zhang
\cite{xingzhang}, for exclusive and inclusive value of $|V_{ub}|$
we have carried out a detailed analysis regarding the structural
features of the mass matrices by incorporating the extended ranges
of the elements $D_i$ $(i = U, D)$ as well as by imposing the
constraints given in equation (\ref{ckmvalues}). To this end, we
first present the results pertaining to exclusive value of
$|V_{ub}|$, the case of its inclusive value will be discussed
later.

\begin{figure}
\psfig{file=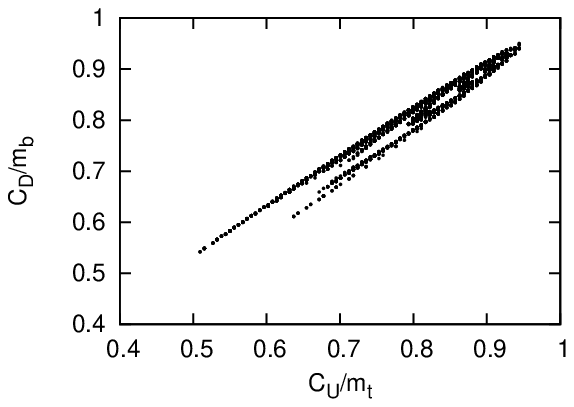, width=3.5in} \caption{Plot showing the
allowed range of $C_{\rm U}/m_t$ versus $C_{\rm D}/m_b$}
  \label{cucd}
  \end{figure}

To begin with, in figure \ref{cucd} we have plotted $C_{U }/m_t$
versus $C_{ D}/m_b$. A look at the figure reveals that both
$C_{U}/m_t$ as well as $C_{ D}/m_b$ take values from $\sim
0.55-0.95$, which interestingly indicates the ratios being almost
proportional. Also, the figure gives interesting clues regarding
the role of strong and weak hierarchy. In particular, one finds
that in case one restricts to the assumption of strong hierarchy
then these ratios take large values around $0.95$. However, for
the case of weak hierarchy, the ratios $C_{U }/m_t$ and $C_{
D}/m_b$ take much larger number of values, in fact almost the
entire range mentioned above, which are compatible with the data.

\begin{figure}
\psfig{file=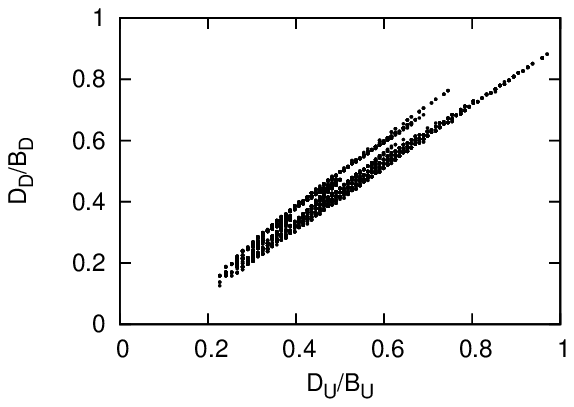, width=3.5in} \caption{Plot showing the
allowed range of $D_{\rm U}/B_{\rm U}$ versus $D_{\rm D}/B_{\rm
D}$}
  \label{dudd}
  \end{figure}

In figure \ref{dudd}, the plot of $D_{U}/B_{U}$ versus $D_{
D}/B_{D}$ has been given which clearly brings out that the ratio
$D_{U}/B_{U} \sim 0.2-0.95$ whereas the ratio $D_{D}/B_{D} \sim
0.15-0.9$. One finds that when the strong hierarchy assumption is
considered then the ratio $D_{U}/B_{U}$ takes value around $0.25$
whereas $D_{D}/B_{D} \sim 0.2$. Again, we find that the ranges of
these ratios corresponding to the weak hierarchy cases are much
wider. The ranges of the parameters plotted in figures \ref{cucd}
and \ref{dudd} not only have implications for the structural
features of the mass matrices, but also indicate that there are
large number of possibilities for which one can achieve
compatibility of texture 4 zero mass matrices with the CKM mixing
data.

\begin{figure}
\psfig{file=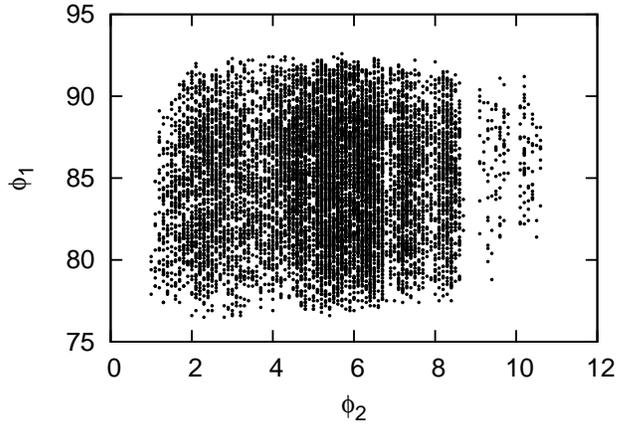, width=3.5in} \caption{Plot showing the
allowed range of $\phi_1$ versus $\phi_2$}
  \label{phi1phi2}
  \end{figure}

In figure \ref{phi1phi2}, we present the plot of $\phi_1$ versus
$\phi_2$. Interestingly, the present refined inputs limit the
ranges of the two phases to $\phi_1 \sim 76^{\rm o} - 92^{\rm o}$
and $\phi_2 \sim 1^{\rm o} - 11^{\rm o}$. Keeping in mind that
full variation has been given to the free parameters $D_U$ and
$D_D$, corresponding to both strong as well as weak hierarchy
cases, it may be noted that the allowed ranges of the two phases
come out to be rather narrow. In particular, for the strong
hierarchy case one gets $\phi_2 \sim 10^{\rm o}$, whereas for the
case of weak hierarchy $\phi_2$ takes almost its entire range
mentioned above. Further, it may be mentioned that this figure
should not be directly compared with the corresponding $\phi_1$
versus $\phi_2$ plot given by \cite{xingzhang} as they have
considered different initial phases. Also, our analysis indicates
that although $\phi_1 \gg \phi_2$, still both the phases are
required for fitting the mixing data. It may also be noted that
the phases $\phi_1$ and $\phi_2$ and the elements of the CKM
mixing matrix can be easily used to obtain angles of the unitarity
triangle.

It may be noted that a comparison of these figures with the
corresponding plots by \cite{xingzhang} immediately reveals that
in the specified ranges of the parameters our results are
compatible with theirs. Further, it may be noted that a direct
comparison of the ranges of various parameters considered by us
and those given by \cite{xingzhang} is not possible, however one
may to able to compare the ranges of the elements of the mass
matrices, which will be discussed later. For the other figures
given in their analysis, it may be mentioned that we obtain the
ones which are quite compatible with their plots as well as lead
to similar consequences and so these are not presented here.

\begin{figure}
\psfig{file=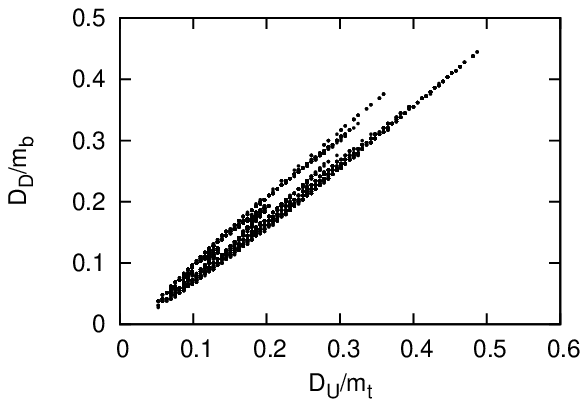, width=3.5in} \caption{Plot showing the
allowed range of $D_{\rm U}/m_t$ versus $D_{\rm D}/m_b$}
  \label{dumtddmb}
  \end{figure}

As a next step, we would like to emphasize the role of the
hierarchy defining parameters $D_{U}$ and $D_{D}$. To this end, in
figure \ref{dumtddmb} we have plotted $D_{U }/m_t$ versus $D_{
D}/m_b$, representing an extended range of the parameters $D_U$
and $D_D$. A closer look at the figure reveals both $D_{U}/m_t$ as
well as $D_{ D}/m_b$ take values $\sim 0.05-0.5$. The lower limit
of the range i.e. when the ratios $D_{U}/m_t$ and $D_{ D}/m_b$ are
around $0.05$ corresponds to strong hierarchy amongst the elements
of the mass matrices, whereas when the elements have weak
hierarchy then these ratios take a much larger range of values.
From this one may conclude that in the case of strongly
hierarchical elements of the texture 4 zero mass matrices, we have
limited compatibility of these matrices with the quark mixing
data, whereas the weakly hierarchical ones indicate the
compatibility for much broader range of the elements.

The above discussion can also be understood by the construction of
the mass matrices. However, as the phases of the elements of the
mass matrices can be separated out, as can be seen from equation
(\ref{mkr}), one needs to consider $M_i^r$ ($i=U, D$) instead of
$M_i$. The ranges of the elements of these matrices $M_U^r$ and
$M_D^r$ are as follows \be M_U^r = m_t \left( \ba {ccc} 0 &
0.000174-0.000252 & 0
\\ 0.000174-0.000252 &0.0464-0.4870 & 0.2184-0.5017
\\0 & 0.2184-0.5017 & 0.5094-0.9500 \ea \right), \label{mu} \ee

\be M_D^r = m_b \left( \ba {ccc} 0 & 0.003555-0.006154 & 0
\\ 0.003555-0.006154 &0.0276-0.4448 & 0.2194-0.5044
\\0 & 0.2194-0.5044 & 0.5418-0.9505 \ea \right). \label{md} \ee
These matrices lead to interesting consequences regarding the
structural features characterized by relative magnitudes of the
elements of the mass matrices. It may be noted that the elements
of the mass matrices $A_i$, $B_i$, $C_i$ and $D_i$ satisfy the
relation $|B_i|^2-C_i D_i \simeq m_2 m_3$ for both the strong and
the weak hierarchy cases characterized respectively by $D_i <
|B_{i}|
< C_i$ and $D_{i} \lesssim |B_i| \lesssim C_i$. This relation can
be easily derived by using expressions mentioned in equation
(\ref{elements}) as well as can be numerically checked from the
above mentioned mass matrices in equations (\ref{mu}) and
(\ref{md}). The above constraint on the elements of the mass
matrices as well as the ranges of various ratios , particularly in
the case of weak hierarchy, provide an interesting possibility for
checking the viability of various mass matrices formulated at the
GUTs scale or obtained using horizontal symmetries. From a
different point of view, this can also provide vital clues to the
formulation of mass matrices which are in agreement with the low
energy data.

Coming to the results pertaining to the inclusive value of
$V_{ub}$, we find that the corresponding figures do not show much
change as compared to the earlier figures plotted using the
exclusive value of $V_{ub}$, therefore these have not been
presented here. The mass matrices $M_U^r$ and $M_D^r$ constructed
using the inclusive value of $V_{ub}$ are as follows
 \be M_U^r = m_t \left( \ba {ccc} 0 &
0.000179-0.000266 & 0
\\ 0.000179-0.000266 &0.0696-0.4928 & 0.2610-0.5018
\\0 & 0.2610-0.5018 & 0.5036-0.9268 \ea \right), \label{muin} \ee

\be M_D^r = m_b \left( \ba {ccc} 0 & 0.003303-0.006462 & 0
\\ 0.003703-0.006462 &0.0552-0.4586 & 0.2686-0.5065
\\0 & 0.2686-0.5065 & 0.5280-0.9236 \ea \right). \label{mdin} \ee
A comparison of these matrices with the ones mentioned in
equations (\ref{mu}) and (\ref{md}) reveals that the (2,2) element
$D_i$ of these matrices appear to be quite different for the
corresponding $M_U^r$ and $M_D^r$ matrices for the case of
exclusive and inclusive values of $V_{ub}$. Similarly, the lower
limits of the element $B_i$ of the mass matrices are quite
different for both the $M_U^r$ and $M_D^r$ matrices corresponding
to exclusive and inclusive values of $V_{ub}$. Therefore, it seems
that refinements in the evaluation of exclusive and inclusive
values of $V_{ub}$ would have implications for the hierarchy of
the elements of the texture 4 zero mass matrices.

As mentioned earlier, it seems interesting to compare the ranges
of the elements of the above mentioned mass matrices with those
constructed by Xing and Zhang \cite{xingzhang}. The comparison
immediately reveals that in the present work we have been able to
achieve agreement with the CKM mixing data for much wider ranges
of the elements of the mass matrices. A closer scrutiny of our
results reveals that these wider ranges are essentially due to
wider ranges for the hierarchy defining parameters $D_U$ and
$D_D$. It may be added that in case we restrict ourselves to the
strong hierarchy case, then we are able to reproduce the matrices
given by \cite{xingzhang}.

After constructing the mass matrices, it is desirable to construct
the corresponding CKM mixing matrix and compare it with the one
arrived through global analysis. To this end, we have considered
the average value of $V_{ub}$ given by PDG 2008, the other input
parameters remain the same. The CKM mixing matrix so obtained is
as follows
 \be V_{{\rm CKM}} = \left( \ba{ccc}
  0.9738-0.9747 &~~~~   0.2236-0.2274 &~~~~  0.00357-0.00429 \\
 0.2234-0.2274  &~~~~   0.9729-0.9739    &~~~~  0.0401-0.0423\\
0.0057-0.0114  &~~~~  0.0388-0.0420 &~~~~  0.9991-0.9992 \ea
\right). \label{1sm} \ee A general look at the matrix reveals that
the ranges of CKM elements obtained here are quite compatible with
those obtained by recent global analyses \cite{pdg08}-
\cite{hfag}. We have also evaluated the Jarlskog's rephasing
invariant parameter $J$ using the average value of $V_{ub}$ which
comes out to be \be J=(1.807 - 3.977) 10^{-5} \label{j}. \ee
Further, using this value of $J$ we obtain the following range of
the CP violating phase $\delta$
 \be \delta=28.8.8^{\rm o}- 110.4^{\rm o}. \label{delta}\ee
The above mentioned ranges of the parameter $J$ and the phase
$\delta$ are inclusive of the values given by PDG 2008
\cite{pdg08}.

\section{Summary and conclusions\label{summ}}
In the light of recent precision measurements, we have made an
attempt to update and broaden the scope of the analysis carried
out by Xing and Zhang \cite{xingzhang} as well as have carried out
a detailed analysis regarding the structural features of the mass
matrices. The implications of these measurements on the texture 4
zero mass matrices have been examined by considering not only the
usual `strong hierarchy' amongst the elements of these matrices,
defined as $D_i < |B_{i}| < C_i$, but also for the `weak
hierarchy' case given by $D_{i} \lesssim |B_i| \lesssim C_i$. For
both the exclusive and inclusive values of $V_{ub}$, the analysis
has been carried out by giving wide variation to the hierarchy
defining parameters $D_U$ and $D_D$. Further, in view of the more
precise information regarding CP violating parameters, the ranges
of both the phases $\phi_1$ and $\phi_2$, having their origin in
the mass matrices, have been found.

We find that despite considering weak hierarchy still both the
phases are required to fit the data, in particular these come out
to be $\phi_1 \sim 76^{\rm o} - 92^{\rm o}$ and $\phi_2 \sim
1^{\rm o} - 11^{\rm o}$. Also for both the exclusive and inclusive
values of $V_{ub}$, the texture 4 zero mass matrices are
compatible with recent results emerging from global fits
\cite{pdg08}- \cite{hfag} for weak as well as strong hierarchy of
the elements of the mass matrices. A comparison of $M_U^r$ and
$M_D^r$ matrices corresponding to exclusive and inclusive values
of $V_{ub}$ reveals that the parameters $D_i$ and $B_i$ ($i=U, D$)
would have implications for these values of $V_{ub}$. In
conclusion, we would like to state that even weakly hierarchical
mass matrices can explain the quark masses and mixing data which
are strongly hierarchical. This, in turn, would have important
implications for model building of the fermion mass matrices.

 \vskip 0.5cm
{\bf Acknowledgements} \\ RV would like to thank the Director,
RIEIT for supporting the work. GA acknowledges DST, Government of
India for financial support and the Chairman, Department of
Physics for providing facilities to work. NM and MR would
respectively like to thank the Chairman, Department of Physics and
the Director, UIET for providing required facilities.

\newpage

\end{document}